\documentclass[a4paper,11pt]{article}

\usepackage{graphicx}

\usepackage{amsmath}
\usepackage{amsfonts}
\usepackage{amssymb}
\usepackage{latexsym}
\usepackage{amsthm}  

\usepackage[textwidth=15cm,textheight=24cm]{geometry}

\begin{document}

\title{Better approximation algorithm for point-set diameter}

\author{Mahdi Imanparast\footnote{Department of Mathematics and Computer science, Amirkabir University of Technology, Tehran, Iran} \hspace*{0.9cm}Seyed Naser Hashemi\footnotemark[1]}

\date{}
\maketitle

\begin{abstract}
We propose a new $(1+O(\varepsilon))$-approximation algorithm with $O(n+ 1/\varepsilon^{\frac{(d-1)}{2}})$ running time for computing the diameter of a set of $n$ points in the $d$-dimensional Euclidean space for a fixed dimension $d$, where $0 < \varepsilon\leqslant 1$. This result provides some improvements in the running time of this problem in comparison with previous algorithms. 

\end{abstract}

\section{Introduction}
For a finite set of $n$ points, the purpose of computing the diameter of a point set is to find two points with maximum distance in the point set. Computing the diameter has a long history and is used in various fields, such as database, data mining, and vision. For example, in a database containing a set of images of the same size, it is possible to consider each image with $d$ pixels as a point in the $d$-dimensional space for a fixed dimension $d$. In this case, computing the diameter of this dataset means finding two most different images. A trivial brute-force solution for the diameter problem is to calculate the distance between each pair of points and then select the maximum distance which takes $O(n^{2})$ time. By reducing from the set disjointness problem, it can be shown that computing the diameter of $n$ points in $\mathbb{R}^{d}$ requires $\Omega(n\log n)$ operations in the algebraic computation-tree model~\cite{Prep85}. There are well-known solutions to this problem in two and three dimensions. This problem can be solved at the optimal time $O(n\log n)$ in the plane, but in three dimensions, it is a bit more difficult. Several attempts have been done to solve the diameter problem at the optimal time in three dimensions, and finally, Ramos~\cite{Ram01} presented an optimal deterministic $O(n \log n)$-time algorithm in $\mathbb{R}^{3}$.
In the case of higher dimensions (for $d>3$), the brute-force algorithm needs $O(dn^{2})$ time, which is too slow for large-scale datasets that appear in the fields. Hence, it seems to be necessary to use approximation algorithms for computing the diameter in these dimensions. A 2-approximation algorithm with $O(dn)$ time in $d$-dimensional space can easily be obtained by choosing a point from the point set, and then finding the farthest point of it by brute-force manner. The first non-trivial approximation algorithm for diameter is presented by Egecioglu and Kalantari~\cite{Ege89}, which calculates a $\sqrt{3}$-approximation with $O(dn)$ time.
They also provided an iterative algorithm with $O(tdn)$ time and an approximation factor $\sqrt{5-2\sqrt{3}}$, where $t$ is the number of iterations of the algorithm. Agarwal et al.~\cite{AMS92} proposed the first $(1+\varepsilon)$-approximation algorithm for computing the diameter in $\mathbb{R}^{d}$ with $O(n/ \varepsilon^{(d-1)/2})$ time by projection to directions. Subsequently, several approximation algorithms have been proposed~\cite{Bar01,HarP01,FP02,MB02,Cha02,Cha06,IHM17,Chan17,AFM17}, each of which improved the required time to solve this problem. Table~\ref{tab:tabe1} provides a summary overview of non-constant approximation algorithms for computing the diameter of a set of points.
 
\vspace {-0.5 cm} 
\begin{table}‎
\caption{A summary of the complexity of some utilizable non-constant approximation algorithms for diameter of a point set in $d$-dimensional Euclidean space. Our result is denoted by +.}‎
\label{tab:tabe1}‎
\vspace {-1.1 cm} 
\begin{center}‎
\begin{tabular}[c]{lccc}
\\
\hline‎
\bf Ref. & \bf  \hspace{0.15cm} Approximation  Factor & \bf \hspace{-0.05cm} Running Time  & \bf \hspace{0.15cm} Year 

\\
\hline‎ 
~\cite{AMS92} & $1+\varepsilon$ &$O(\dfrac{n^{^{}}}{\varepsilon^{(d-1)/2}})$ &1992  \\
\hspace*{0.1cm}~\cite{Bar01} & $1+\varepsilon$   &$O(n+1/\varepsilon^{2(d-1)})$ & 2001  \\‎
~\cite{HarP01}  & $1+\varepsilon$ & $O((n+ 1/\varepsilon^{2d}) \log \frac{1}{\varepsilon})$ & 2001  \\
\hspace*{0.1cm}~\cite{Cha02}  & $1+\varepsilon$ &  $O(n+1/\varepsilon^{\frac{3(d-1)}{2}})$  & 2002  \\‎
~\cite{IHM17}   & $1+\varepsilon$  & $O(n+ 1/\varepsilon^{d-2})$  & 2018\\
\hspace*{0.1cm}~\cite{Cha02}  & $1+O(\varepsilon)$  & $O(n+1/\varepsilon^{d-\frac{1}{2}})$   & 2002 \\‎
~\cite{Cha06}  & $1+O(\varepsilon)$   & $O(n+1/\varepsilon^{d-\frac{3}{2}})$ & 2006 
\\‎

~\cite{IHM17} & $1+O(\varepsilon)$  & $O(n+ 1/\varepsilon^{\frac{2d}{3}-\frac{1}{2}})$ & 2018

\\‎
~\cite{Chan17}   & $1+O(\varepsilon)$  & $O((n/\sqrt{\varepsilon}+1/\varepsilon^{\frac{d}{2}+1}) (\log\frac{1}{\varepsilon})^{O(1)})$  & 2017

\\‎
~\cite{AFM17}  & $1+O(\varepsilon)$ & $O(n \log \frac{1}{\varepsilon}+1/\varepsilon^{\frac{(d-1)}{2}+\alpha})$   & 2017
\\
\hspace*{0.27cm} +  & $1+O(\varepsilon)$  & $O(n+ 1/\varepsilon^{\frac{(d-1)}{2}})$ & 2018
\\
\hline‎ 
\end{tabular}‎
\end{center}‎
\end{table}‎

\vspace {-0.12 cm}  

\subsection{Our result}
In this paper, we propose a simple $(1+O(\varepsilon))$-approximation algorithm for computing the diameter of a set $\mathcal{S}$ of $n$ points in $\mathbb{R}^{d}$ with $O(n+ 1/\varepsilon^{\frac{(d-1)}{2}})$ time and $O(n)$ space, where $0 < \varepsilon\leqslant 1$. As stated in Table~\ref{tab:tabe1}, two new results have been recently presented for the diameter problem in~\cite{Chan17} and~\cite{AFM17}. It should be noted that our algorithm is completely different in terms of computational technique. The polynomial technique provided by Chan~\cite{Chan17} is based on using Chebyshev polynomials and discrete
upper envelope subroutine~\cite{Cha06}, and the method presented by Arya et al.~\cite{AFM17} requires the use of complex data structures to approximately answer
queries for polytope membership, directional width, and nearest-neighbor. While our algorithm in comparison with these algorithms is simpler in terms of understanding and data structure. 
\section{The proposed algorithm}	
In this section, we describe our new approximation algorithm to compute the diameter of a set $\mathcal{S}$ of $n$ points in $\mathbb{R}^{d}$. The main idea of our algorithm is based on the rounding points to the grids. We first round the points to their central-cell points of the grid and then in two phases to their nearest grid-points for different grid cell sizes. The most important advantage of these sequential rounding by increasing the grid cell size is that it will not only reduce the number of points we examine, but also it will reduce the search domain for computing  the diameter at each rounding phase. In fact, by computing the diameter in the set of rounded points in the third phase, we divide the problem into a set of subproblems that helps us in solving subproblems in the previous rounded point sets and ultimately leads to finding a $(1+O(\varepsilon))$-approximation of the true diameter. 
\vspace {-0.2 cm} 
\begin{figure}
    \begin{center}
       \includegraphics [height=3.5 cm]{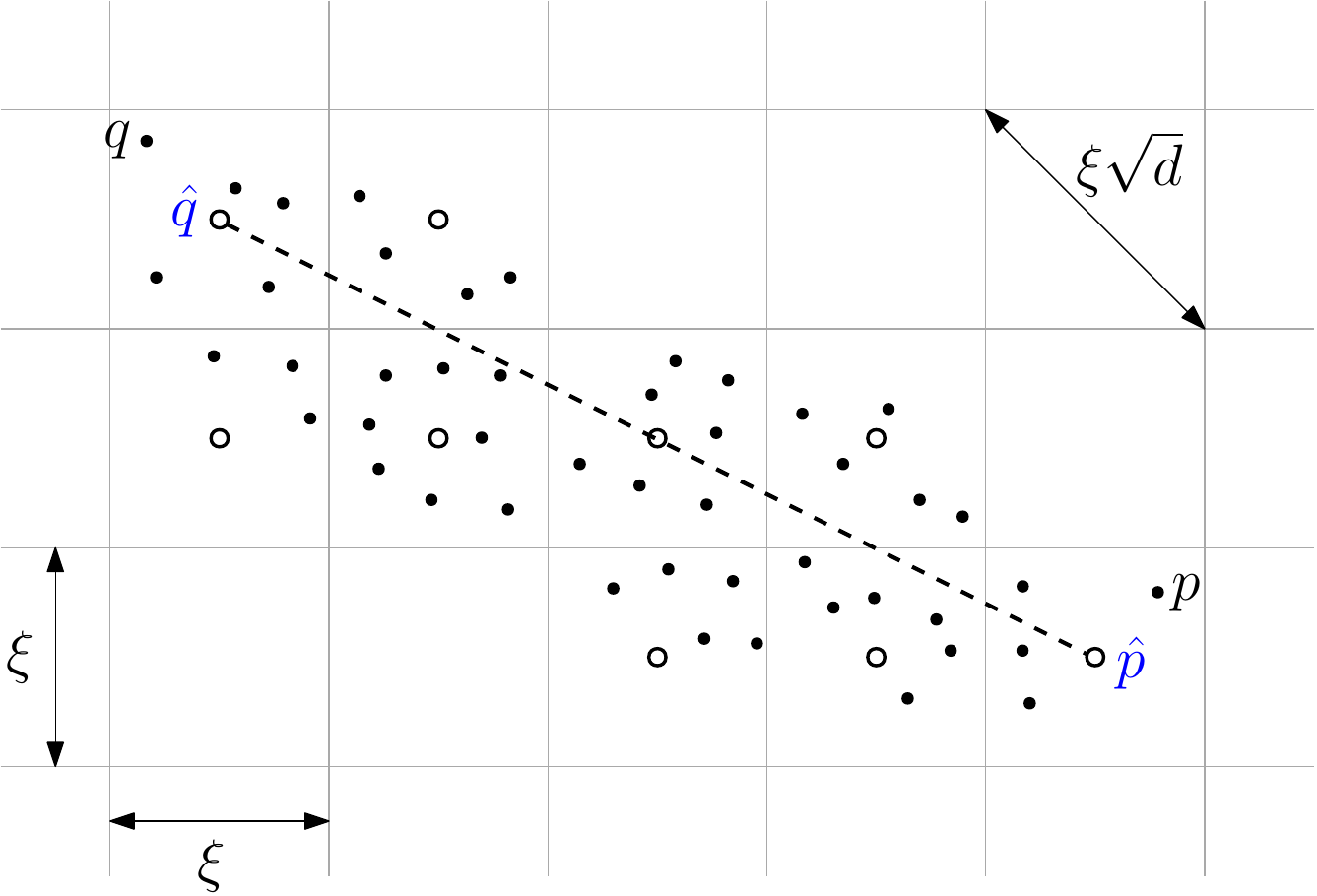}
       \caption{Rounding the points of the set $\mathcal{S}$ to their central-cell points in a $\xi$-grid. Two points $\hat{p}$ and $\hat{q}$ are the corresponding central-cell points for two points $p$ and $q$ which are the diametrical pair $(p,q) \in \mathcal{S}$.}
       \vspace {-0.2 cm}
      \label{fig:Fig-1}                   
    \end{center}
\end{figure}
\vspace {0.2 cm}

We first find the extreme points in each coordinate and compute the axis-parallel bounding box of $\mathcal{S}$, which is denoted by $B(\mathcal{S})$. We use the largest length side $\ell$ of $B(\mathcal{S})$ to impose grids on the point set. In first rounding phase, we decompose $B(\mathcal{S})$ to a grid of regular hypercubes with side length $\xi$, where $\xi=\varepsilon \ell/2\sqrt{d}$. We call each hypercube a cell. See Fig.~\ref{fig:Fig-1}. This rounding provides a $(1+\varepsilon)$-approximation for the diametrical pair $(p,q) \in \mathcal{S}$ (see details in~\cite{IHM17}). We assume that $\mathcal{B}_{\delta}(p)$ denotes a hypercube with side length $\delta$ and central-point $p$. We also use notation $Diam(\mathcal{B}_{1},\mathcal{B}_{2})$ for the process of computing the diameter of the point set $\mathcal{B}_{1}\cup \mathcal{B}_{2}$. In the following, we present our algorithm in more details in Algorithm 1.

\vspace*{0.2cm}

\smallskip‎
\hrule‎
\\
\textbf{ Algorithm 1‎:} APPROXIMATE DIAMETER $ (\mathcal{S},\varepsilon)$ 
\hrule‎
\\ 
\footnotesize‎ 
\smallskip‎
\textbf{‎Input‎:} a set $\mathcal{S}$ of $n$ points in $\mathbb{R}^{d}$ and an error parameter $\varepsilon$.

\hspace*{-0.4cm}\textbf{‎Output‎:} approximate diameter $\hat{D}$.

\hspace{-0.4 cm} 1: \hspace{0.03cm} Compute the axis-parallel bounding box $B(\mathcal{S})$ for the point set $\mathcal{S}$.

\hspace{-0.4 cm} 2: \hspace{0.03cm} $\ell \leftarrow$ Find the length of the largest side in $B(\mathcal{S})$. 

\hspace{-0.4 cm} 3: \hspace{0.03cm} Set $\xi\leftarrow \varepsilon \ell/2\sqrt{d}$,  $\xi_{1}\leftarrow \sqrt{\varepsilon} \ell/2\sqrt{d}$ and $\xi_{2}\leftarrow \varepsilon^{\frac{1}{4}} \ell/2\sqrt{d}$.

\hspace{-0.4 cm} 4: \hspace{0.03cm} $\hat{\mathcal{S}}\leftarrow$ Round each point of $\mathcal{S}$ to its central-cell point in a $\xi$-grid.

\hspace{-0.4 cm} 5: \hspace{0.03cm} $\hat{\mathcal{S}'}\leftarrow$ Round each point of $\hat{\mathcal{S}}$ to its nearest grid-point in a $\xi_{1}$-grid.

\hspace{-0.4 cm} 6: \hspace{0.03cm} $\hat{\mathcal{S}''}\leftarrow$ Round each point of $\hat{\mathcal{S}'}$ to its nearest grid-point in a $\xi_{2}$-grid.

\hspace{-0.4 cm} 7: \hspace{0.03cm} $\hat{D}'' \leftarrow$ Compute the diameter of the point set $\hat{\mathcal{S}''}$ by brute-force manner.

\hspace{-0.4 cm} 8: \hspace{0.06cm} Compute a list of the diametrical pairs $(\hat{p}'',\hat{q}'')$, such that $\hat{D}''=||\hat{p}''-\hat{q}''||$.

\hspace{-0.4 cm} 9: \hspace{0.06cm} Corresponding to each diametrical pair $(\hat{p}'',\hat{q}'') \in \hat{\mathcal{S}''}$, find points of $\hat{\mathcal{S}'}$ which \\ \hspace*{0.86cm} are inside two hypercubes $\mathcal{B}_{2\xi_{2}}(\hat{p}'')$ and $\mathcal{B}_{2\xi_{2}}(\hat{q}'')$, and store them in two point\\ \hspace*{0.86cm} sets $\mathcal{B}_{1}'$  and $\mathcal{B}_{2}'$. 

\hspace{-0.4 cm} 10: \hspace{0.04cm} $\hat{D}'\leftarrow$ For each pair $(\mathcal{B}_{1}', \mathcal{B}_{2}')$ corresponding to each diametrical pair $(\hat{p}'',\hat{q}'') \in \hat{\mathcal{S}''}$,\\ \hspace*{1.7cm} compute $Diam(\mathcal{B}_{1}',\mathcal{B}_{2}')$, by brute-force manner and return the maximum \\ \hspace*{1.7cm} value  between them.

\hspace{-0.4 cm} 11: \hspace{0.04cm} Compute a list of the diametrical pairs $(\hat{p}',\hat{q}')$, such that $\hat{D}'=||\hat{p}'-\hat{q}'||$.

\hspace{-0.4 cm} 12: \hspace{0.04cm} Corresponding to each diametrical pair $(\hat{p}',\hat{q}') \in \hat{\mathcal{S}'}$, find points of $\hat{\mathcal{S}}$ which \\ \hspace*{0.95cm}  are inside two hypercubes $\mathcal{B}_{2\xi_{1}}(\hat{p}')$ and $\mathcal{B}_{2\xi_{1}}(\hat{q}')$, and store them in two point\\ \hspace*{0.98cm} sets $\mathcal{B}_{1}$  and $\mathcal{B}_{2}$. 

\hspace{-0.4 cm} 13: \hspace{0.04cm} $\hat{D}\leftarrow$ Compute $Diam(\mathcal{B}_{1},\mathcal{B}_{2})$, corresponding to each diametrical pair $(\hat{p}',\hat{q}')\in \hat{\mathcal{S}'}$ \\ \hspace*{1.75cm} by Chan's~\cite{Cha02} recursive approach and return the maximum value between\\ \hspace*{1.75cm} them.

\hspace{-0.4 cm} 14: \hspace{0.01cm} Output $\hat{D}$.
\smallskip‎
\hrule‎
\rmfamily‎ 
\normalsize‎

\vspace {-0.4 cm} 
\begin{figure}
    \begin{center}
       \includegraphics [height=8.75 cm]{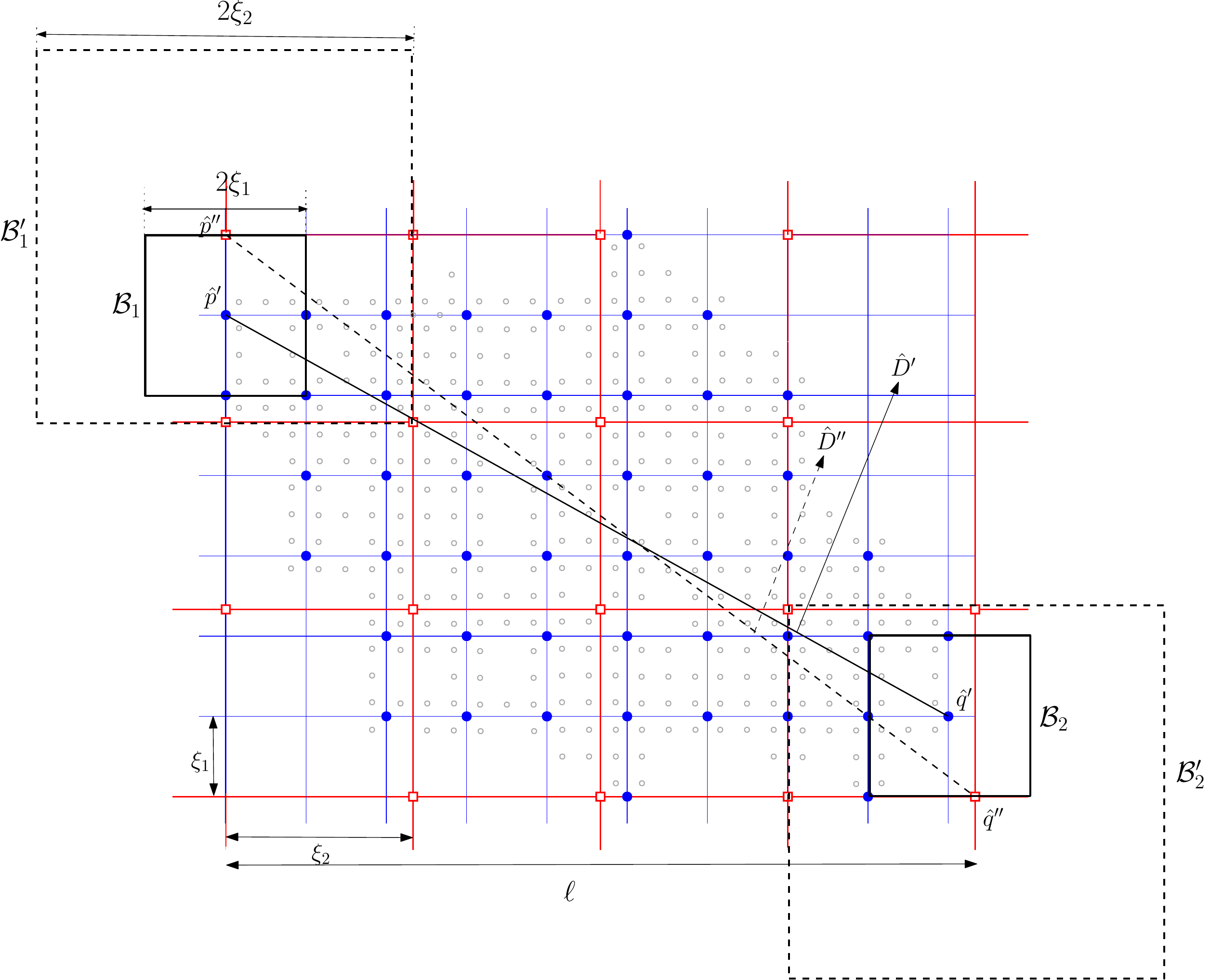}
       \caption{An example of rounding the points of the rounded point set $\hat{\mathcal{S}}$ to a $\xi_{1}$-grid, and then rounding the points of the set $\hat{\mathcal{S}'}$ to a $\xi_{2}$-grid. }
       \vspace {-0.2 cm}
      \label{fig:Fig-2}                   
    \end{center}
\end{figure}
\vspace {0.52 cm}

In Fig.~\ref{fig:Fig-2}, we illustrate an example of rounding the points of the rounding point set $\hat{\mathcal{S}}$ to a $\xi_{1}$-grid, and then rounding points of the new point set $\hat{\mathcal{S}'}$ to a $\xi_{2}$-grid. Points of the rounded point set $\hat{\mathcal{S}}$ are shown by circle points ($\circ$) and their corresponding nearest grid-points in point set $\hat{\mathcal{S}'}$ are shown by blue points ($\bullet$). Moreover, points of the point set $\hat{\mathcal{S}''}$ are shown by small red squares. The searching domain for finding the diameter of the point set $\hat{\mathcal{S}}$ is reduced into two point sets $\mathcal{B}_{1}$ and $\mathcal{B}_{2}$. It should be noted that we use a hypercube $\mathcal{B}_{2\xi_{1}}(\hat{p}')$ of side length $2\xi_{1}$ to make sure that we do not lose any points of the rounded point set $\hat{\mathcal{S}}$ around a grid-point $\hat{p}'\in \hat{\mathcal{S}'}$.

What needs to be further explained is the last step of the algorithm in which Chan's~\cite{Cha02} recursive approach is used to compute the approximate diameter. As previously mentioned, Agarwal et al.~\cite{AMS92} introduced an approximation algorithm that in $O(n/\varepsilon^{(d-1)/2})$ time could find a $(1+\varepsilon)$-approximation for the diameter of a set of $n$ points. Their result is based on the following simple observation. Let denote the projection of a point set $\mathcal{S}$ on the line $l$ by $\mathcal{S}_{l}$. Then, if two points $p$ and $q$ are the diametrical pair of the point set $\mathcal{S}$, and $p'$ and $q'$ be their projection on the line $l$, such that the angle between $pq$ and $p'q'$ be at most $\sqrt{\varepsilon}$, we have $||p'-q'||\leqslant||p-q||\leqslant (1+\varepsilon)||p'-q'||$. This means that $||p'-q'||$ is a $(1+\varepsilon)$-approximate of $||p-q||$. Now we can find $O(1/\varepsilon^{(d-1)/2})$ number of directions in $\mathbb{R}^{d}$, for example by constructing a uniform grid on a unit sphere, such that for each vector $x \in\mathbb{R}^{d}$, there is a direction that the angle between this direction and vector $x$ be at most $\sqrt{\varepsilon}$. So, it is sufficient to project the point set $\mathcal{S}$ on each of these directions and compute their diameter. Then, we can consider the maximum value of the computed diameters as a $(1+\varepsilon)$-approximate diameter of the point set $\mathcal{S}$. Chan~\cite{Cha02} observed that instead of running their method on $n$ points, the points can be first rounded to a grid, in which case the number of points would be reduced from $n$ to $m=O(1/\varepsilon^{(d-1)})$. Then, by applying Agarwal et al.~\cite{AMS92} method on this rounded point set, we need $O((1/\varepsilon^{d-1})/\varepsilon^{(d-1)/2})=O(1/\varepsilon^{3(d-1)/2})$ time to compute the maximum diameter over all $O(1/\varepsilon^{(d-1)/2})$ directions. Taking into account $O(n)$ time we spend for rounding to a grid, this new approach computes a $(1+\varepsilon)$-approximation for the diameter of a set of $n$ points in $O(n+1/\varepsilon^{3(d-1)/2})$ time.

Chan also observed that in Agarwal et al. method, instead of projecting rounded points on $O(1/\varepsilon^{(d-1)/2})$ $d$-dimensional unit vectors, one can project $m=O(1/\varepsilon^{(d-1)})$ rounded points on a set of $O(1/\sqrt{\varepsilon})$ 2-dimensional unit vectors to reduce the problem to $O(1/\sqrt{\varepsilon})$ numbers of $(d-1)$-dimensional subproblems which can be solved recursively. Let us denote the required time for computing the diameter of $m$ points in a $d$-dimensional space with $t_{d}(m)$, then for a rounded point set on a grid with $m=O(1/\varepsilon^{d-1})$ points, this recursive approach breaks the
problem into $O(1/\sqrt{\varepsilon})$
subproblems in a $(d-1)$ dimensional space. Hence, we have a recurrence 
\begin{equation}   
t_{d}(m)=O(m+1/\sqrt{\varepsilon}t_{d-1}(O(1/\varepsilon^{d-1}))). 
\end{equation} 
By assuming $E = 1/\varepsilon$, we can rewrite the recurrence as:
\begin{equation}
t_{d}(m)=O(m+E^{\frac{1}{2}}t_{d-1}(O(E^{d-1}))).
\end{equation} 
This can be solved to $t_{d}(m)=O(m+E^{d-\frac{1}{2}})$.
In this case, $m=O(1/\varepsilon^{d-1})$, so, this recursion takes $O(1/\varepsilon^{d-\frac{1}{2}})$ time. Taking into account $O(n)$ time, we spent for rounding to a grid at the first, Chan's recursive approach computes a $(1+O(\varepsilon))$-approximation for the diameter of a set of $n$ points in $O(n+1/\varepsilon^{d-\frac{1}{2}})$ time~\cite{Cha02}. For more details on the Agarwal et al. and Chan's recursive methods, the readers are referred to~\cite{Cha02} and~\cite{IHM17}.

\subsection{Analysis}
In this subsection, we analyze the proposed algorithm.

\newtheorem{The1} [enumi] {Theorem}{\bfseries}{\itshape}
\begin{The1}
Algorithm 1 computes a $(1+O(\varepsilon))$-approximate diameter for a set $\mathcal{S}$ of $n$ points in $\mathbb{R}^{d}$ in $O(n+1/\varepsilon^{\frac{(d-1)}{2}})$ time and $O(n)$ space, where $0<\varepsilon\leqslant 1$.
\end{The1}
\begin{proof}
Finding the extreme points in all coordinates and finding the largest side of $B(\mathcal{S})$ can be done in $O(dn)$ time. The rounding step takes $O(d)$ time for each point, and for all of them takes $O(dn)$ time. But for computing the diameter over the rounded point set $\hat{\mathcal{S}''}$, in line 7 of the Algorithm 1, we need to know the number of points in the set $\hat{\mathcal{S}''}$. We know that the largest side of the bounding box $B(\mathcal{S})$ has length $\ell$ and the side length of each cell in the $\xi_{2}$-grid is $\xi_{2}=\varepsilon^{\frac{1}{4}}\ell/2\sqrt{d}$. On the other hand, the volume of a hypercube of side length $L$ in $d$-dimensional space is $L^{d}$. Since, corresponding to each point in the point set $\hat{\mathcal{S}''}$, we can take a hypercube of side length $\xi_{2}$. Therefore, in order to count the maximum number of points inside the point set $\hat{\mathcal{S}''}$, it is sufficient to calculate the number of hypercubes of side length $\xi_{2}$ in a hypercube (bounding box) with side length $\ell+\xi_{2}$. See Fig.~\ref{fig:Fig-2}. This means that the number of grid-points in an imposed $\xi_{2}$-grid to the bounding box $B(\mathcal{S})$ is at most
\begin{equation}
\dfrac{(\ell+\xi_{2})^{d}}{(\xi_{2})^{d}}=(\dfrac{\ell}{\varepsilon^{\frac{1}{4}}\ell/2\sqrt{d}}+1)^{d}=(\dfrac{2\sqrt{d}} {\varepsilon^{\frac{1}{4}}}+1)^{d}
=O(\dfrac{(2\sqrt{d})^{d}}{\varepsilon^{\frac{d}{4}}}) .
\end{equation}
So, the number of points in $\hat{\mathcal{S}''}$ is at most $O((2\sqrt{d})^{d} / \varepsilon^{\frac{d}{4}})$. This can be reduced to $O((2\sqrt{d})^{d} / \varepsilon^{\frac{d}{4}-1})$ by discarding some internal points which do not have any potential to be the diametrical pairs in rounded point set $\hat{\mathcal{S}''}$. This can be done by considering all the points which are same in their $(d-1)$ coordinates and keep only highest and lowest. Hence, by the brute-force quadratic algorithm, we need $O((2\sqrt{d})^{d} / \varepsilon^{\frac{d}{4}-1})^{2})=O((2\sqrt{d})^{2d} / \varepsilon^{\frac{d}{2}-2})$ time for computing all distances between points of the set $\hat{\mathcal{S}''}$, and its diametrical pair list. Then, for a diametrical pair $(\hat{p}'',\hat{q}'')$ in point set $\hat{\mathcal{S}''}$, we compute two sets $\mathcal{B}_{1}'$ and $\mathcal{B}_{2}'$ in  $O(dn)$ time. They include points of $\hat{\mathcal{S}'}$ which are inside two hypercubes $\mathcal{B}_{2\xi_{1}}(\hat{p}'')$ and $\mathcal{B}_{2\xi_{1}}(\hat{q}'')$, respectively. In addition, for computing the diameter of the point set $\mathcal{B}_{1}'\cup \mathcal{B}_{2}'$, we need to know the number of points in each of two point sets $\mathcal{B}_{1}'$ and $\mathcal{B}_{2}'$. On the other hand, the number of points in two sets $\mathcal{B}_{1}'$ or $\mathcal{B}_{2}'$ is equivalent to the number of hypercubes of size $\mathcal{B}_{\xi_{1}}$ that can be in a hypercube of size $\mathcal{B}_{2\xi_{2}}$, which is at most
\begin{equation}
\dfrac{Vol(\mathcal{B}_{2\xi{2}})}{Vol(\mathcal{B}_{\xi_{1}})}=\dfrac{(2\varepsilon^{\frac{1}{4}} \ell/ 2\sqrt{d})^{d}}{(\sqrt{\varepsilon} \ell/ 2\sqrt{d})^{d}}=\dfrac{(2 \varepsilon^{\frac{1}{4}})^{d}}{(\varepsilon^{\frac{1}{2}})^{d}}
=\dfrac{(2)^{d}}{\varepsilon^{\frac{d}{4}}}.
\end{equation}

\vspace {-0.4 cm} 
\begin{figure}
    \begin{center}
       \includegraphics [height=4.95 cm]{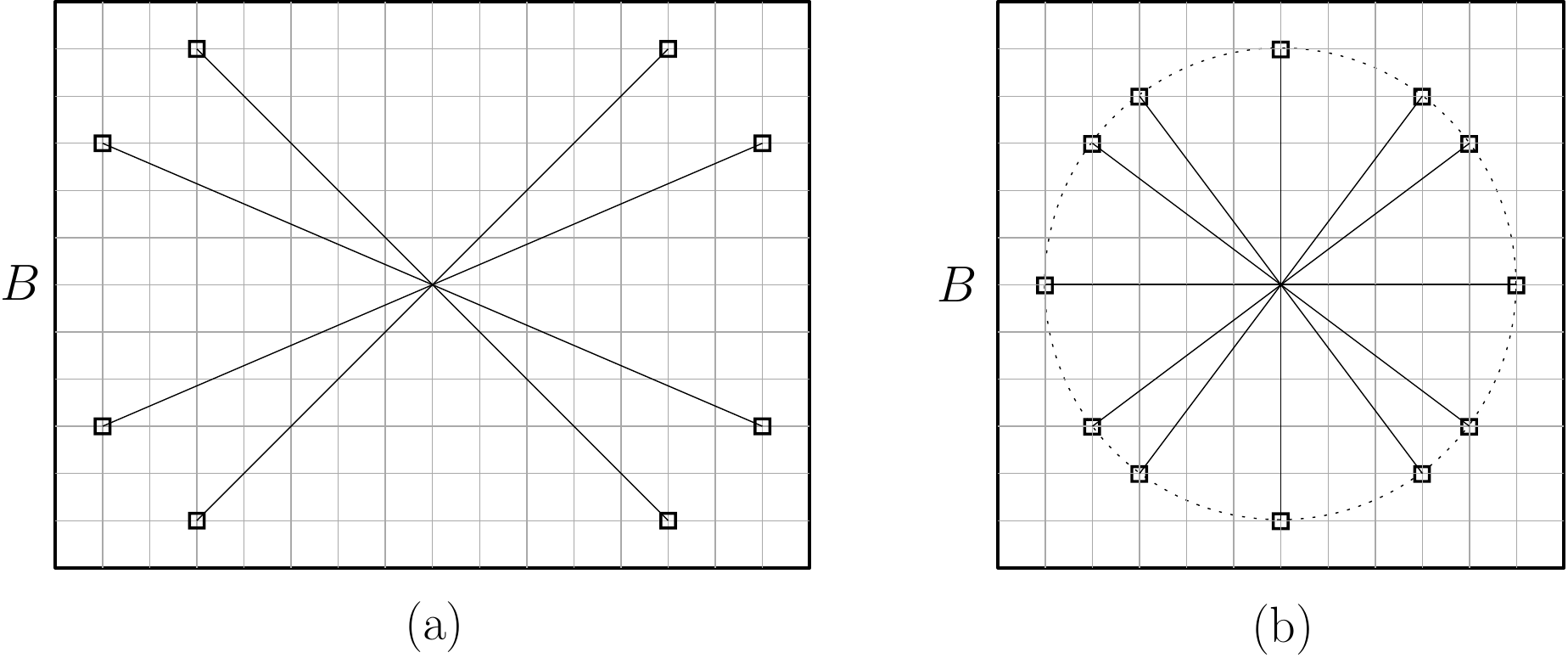}
       \caption{The maximum number of diametrical pairs on a regular grid in two dimensional space. The squares ($\Box$) denote the rounded points on the grid. }
       \vspace {-0.2 cm}
      \label{fig:Fig-3}                   
    \end{center}
\end{figure}
\vspace {0.52 cm}

This means that the number of points in two point sets $\mathcal{B}_{1}'$ and $\mathcal{B}_{2}'$ is at most $O((2)^{d}/\varepsilon^{\frac{d}{4}})$, which can be also reduced to $O(2^{d}/\varepsilon^{\frac{d}{4}-1})$. Hence, for computing $Diam(\mathcal{B}_{1}',\mathcal{B}_{2}')$, we need $O(((2)^{d}/\varepsilon^{\frac{d}{4}-1})^{2})=O((2)^{2d}/\varepsilon^{\frac{d}{2}-2})$ time by brute-force manner. But we might have more than one diametrical pair $(\mathcal{B}_{1}',\mathcal{B}_{2}')$. Since the point set $\hat{\mathcal{S}}''$ is a set of grid-points, so we could have in the worst-case $O(2^{d})$ different diametrical pairs $(\mathcal{B}_{1}',\mathcal{B}_{2}')$ in the point set $\hat{\mathcal{S}}''$. Two examples of the diametrical pairs on regular girds in the two-dimensional space are shown in Fig.~\ref{fig:Fig-3}. Therefore, this step takes at most $O(2^{d}\cdot (2)^{2d}/\varepsilon^{\frac{d}{2}-2})= O((2\sqrt{2})^{2d}/\varepsilon^{\frac{d}{2}-2})$ time. 

In the next step, for each diametrical pair $(\hat{p}',\hat{q}')\in\hat{\mathcal{S}'}$, we compute two sets $\mathcal{B}_{1}$ and $\mathcal{B}_{2}$ which include points of set $\hat{\mathcal{S}}$ which are inside two hypercubes $\mathcal{B}_{2\xi{1}}(\hat{p}')$ and $\mathcal{B}_{2\xi{1}}(\hat{q}')$, respectively. Moreover, the number of points in two sets $\mathcal{B}_{1}$ or $\mathcal{B}_{2}$ is at most
\begin{equation}
\dfrac{Vol(\mathcal{B}_{2\xi{1}})}{Vol(\mathcal{B}_{\xi})}=\dfrac{(2\sqrt{\varepsilon} \ell/ 2\sqrt{d})^{d}}{(\varepsilon \ell/ 2\sqrt{d})^{d}}=\dfrac{(2\varepsilon^{\frac{1}{2}})^{d}} {\varepsilon^{d}}
=\dfrac{(2)^{d}}{\varepsilon^{\frac{d}{2}}}.
\end{equation}
This can be reduced to $O((2)^{d}/\varepsilon^{\frac{d}{2}-1})$, by keeping only highest and lowest points which are the same in their $(d - 1)$ coordinates. Now, for computing $Diam(\mathcal{B}_{1},\mathcal{B}_{2})$, we use Chan's~\cite{Cha02} recursive approach instead of using the quadratic brute-force algorithm on the point set $\mathcal{B}_{1}\cup \mathcal{B}_{2}$. On the other hand, computing the diameter on a set of $O(1/\varepsilon^{\frac{d}{2}-1})$ points using Chan's recursive approach takes the following recurrence based on the relation (1): $t_{d}(m)=O(m+1/\sqrt{\varepsilon}t_{d-1}(O(1/\varepsilon^{\frac{d}{2}-1})))$.
By assuming $E = 1/\varepsilon$, we
can rewrite the recurrence as:
\begin{equation}
t_{d}(m)=O(m+E^{\frac{1}{2}}t_{d-1}(O(E^{\frac{d}{2}-1}))).
\end{equation} 
This can be solved to $t_{d}(m)=O(m+E^{\frac{d}{2}-\frac{1}{2}})$. In this case, $m=O(E^{\frac{d}{2}-1})$, so, this recursion takes $O(E^{\frac{d}{2}-\frac{1}{2}})=O(1/\varepsilon^{\frac{d}{2}-\frac{1}{2}})$ time. Moreover, if we have more than one diametrical pair $(\hat{p}',\hat{q}')$ in point set $\hat{\mathcal{S}'}$, then this step takes at most $O( (2^{d})(2)^{d}/\varepsilon^{\frac{d}{2}-\frac{1}{2}})=O( 2^{2d}/\varepsilon^{\frac{d}{2}-\frac{1}{2}})$ time. Therefore, we can write the total complexity time of the algorithm as follows, which includes required time for three times rounding the point sets to the  grids with $O(dn)$ time, and the required time to find points inside the hypercubes $(\mathcal{B}_{1},\mathcal{B}_{2})$ or $(\mathcal{B}_{1}',\mathcal{B}_{2}')$, which can take at most  $O(2^{d}dn)$ time, plus the time of calculating the diameter in each of the three sets of rounded points: 
$$\hspace{1.9cm}T_{d}(n)=O(dn)+O(dn)+O(dn)+O(2^{d}dn)+O(\dfrac{ (2\sqrt{d})^{2d}}{\varepsilon^{\frac{d}{2}-2}})$$
$$\hspace{0.8cm}+ O(2^{d}dn)+ O(\dfrac{ (2\sqrt{2})^{2d}}{\varepsilon^{\frac{d}{2}-2}})+O(\dfrac{2^{2d}}{\varepsilon^{\frac{d}{2}-\frac{1}{2}}}),$$
\begin{equation}
\hspace{-1.5cm}\leqslant O(2^{d}dn+ \dfrac{(2\sqrt{d})^{2d}}{\varepsilon^{\frac{d}{2}-\frac{1}{2}}}).
\end{equation}
Since $d$ is fixed, we have:
\begin{equation}
\hspace{-2.15cm} T_{d}(n)=O(n+\dfrac{1}{\varepsilon^{\frac{(d-1)}{2}}}).
\end{equation}
On the other hand, we know that the points of the set $\hat{\mathcal{S}}$ provide a $(1+\varepsilon)$-approximate for the true diameter of the point set $\mathcal{S}$ (see Fig.~\ref{fig:Fig-1}), and also the Chan's recursive approach computes a $(1+O(\varepsilon))$-approximate for the diameter of the point set $\hat{\mathcal{S}}$. So, the proposed algorithm provides a $(1+O(\varepsilon))$-approximation diameter of the initial point set $\mathcal{S}$. About the required space, we only need $O(n)$ space for storing required point sets. So, this completes the proof.
\end{proof}
\section{Conclusion}
In this paper, we have presented a simple $(1+O(\varepsilon))$-approximation algorithm to compute the diameter of a point set $\mathcal{S}$ of $n$ points in $\mathbb{R}^{d}$ for a fixed dimension $d$ with $O(n+ 1/\varepsilon^{\frac{(d-1)}{2}})$ time, where $0 <\varepsilon\leqslant 1$.

\end{document}